# On the Bloch Theorem and Orthogonality Relations

Sina Khorasani
*School of Electrical Engineering*
*Sharif University of Technology*
*P. O. Box 11365-9363*
*Tehran, Iran*
Email: khorasani@sina.sharif.edu

**Abstract**
Bloch theorem for a periodic operator is being revisited here, and we notice extra orthogonality relationships. It is shown that solutions are bi-periodic, in the sense that eigenfunctions are periodic with respect to one argument, and pseudo-periodic with respect to the other. An additional kind of symmetry between **r**-space and **k**-space exists between the envelope and eignfunctions not apparently noticed before, which allows to define new invertible modified Wannier functions. As opposed to the Wannier functions in **r**-space, these modified Wannier functions are defined in the **k**-spaces, but satisfy similar basic properties. These could result in new algorithms and other novel applications in the computational tools of periodic structures.

## 1. Introduction

Consider a self-adjoint linear operator problem in the real physical space, denoted by the **r**-space, as

$$\mathcal{L}A_\lambda(\mathbf{r}) = \lambda A_\lambda(\mathbf{r})$$

where $\mathcal{L} = \mathcal{L}^\dagger$, $\lambda$ are eigenvalues and $A_\lambda(\mathbf{r})$ are eigenfunctions. We define a translation operator $\mathcal{T}_\mathbf{R}$ [1] as

$$\mathcal{T}_\mathbf{R} f(\mathbf{r}) = f(\mathbf{r} + \mathbf{R})$$

Now, we take the basis vectors $\mathbf{a}_n$, $n = 1,2,3$, with the non-zero triple product $V = \mathbf{a}_1 \cdot \mathbf{a}_2 \times \mathbf{a}_3$. We may also define the discrete vectors $\mathbf{R} = n_1\mathbf{a}_1 + n_2\mathbf{a}_2 + n_3\mathbf{a}_3$ where $n_1$, $n_2$, and $n_3$ are integers. The set of all such discrete vectors are referred to as the lattice sites [2-4]. Now, suppose that these two operators commute

$$[\mathcal{L}, \mathcal{T}_\mathbf{R}] = 0$$

Then these two operators share the same set of eigenfunctions. Let $\mathcal{T}_\mathbf{R}$ satisfy the eigenvalue equation

$$\mathcal{T}_\mathbf{R} A_\lambda(\mathbf{r}) = \mu A_\lambda(\mathbf{r})$$

The relationship between the sets of eigenvalues $\mu$ and $\lambda$ is generally complicated and highly nontrivial [5].

We may now define a set of reciprocal lattice vectors $\mathbf{b}_1 = \frac{2\pi}{V}\mathbf{a}_2 \times \mathbf{a}_3$, $\mathbf{b}_2 = \frac{2\pi}{V}\mathbf{a}_3 \times \mathbf{a}_1$, and $\mathbf{b}_3 = \frac{2\pi}{V}\mathbf{a}_1 \times \mathbf{a}_2$ with the triple product $U = \mathbf{b}_1 \cdot \mathbf{b}_2 \times \mathbf{b}_3 = (2\pi)^3/V$ [2-4].

Defining a reciprocal lattice vector $\mathbf{K} = m_1 \mathbf{b}_1 + m_2 \mathbf{b}_2 + m_3 \mathbf{b}_3$ we may call the set of all such discrete reciprocal lattice vectors as the reciprocal lattice points. For any choice of the discrete vectors $\mathbf{R}$ and $\mathbf{K}$ we now get the fundamental relationship

$$e^{j\mathbf{K}\cdot\mathbf{R}} = 1$$

For any given discrete lattice vector $\mathbf{R}$, this relationship allows infinite discrete solutions for $\mathbf{K}$ at the reciprocal lattice vectors, and vice versa. In a three-dimensional space, this will dictate a three-fold degeneracy on the eigenvalues and eigenfunctions, albeit in the form of discrete translational symmetry to be discussed later below.

## 2. Bloch Theorem

The basis of the well-known Bloch-Floquet theorem [2-7] is that by which the eigenfunctions of $\mathcal{L}$ take on the property

$$A(\mathbf{r} + \mathbf{R}; \mathbf{k}) = e^{-j\mathbf{R}\cdot\mathbf{k}} A(\mathbf{r}; \mathbf{k})$$

where $\mathbf{k}$ is defined as the Bloch wavevector, and correspondingly, the set of all such vectors is referred to as the $\mathbf{k}$-space. In other words, $A(\mathbf{r}; \mathbf{k})$ is pseudo-periodic in the $\mathbf{r}$-space. Accordingly, the eigenfunctions satisfy

$$A(\mathbf{r}; \mathbf{k}) = e^{-j\mathbf{r}\cdot\mathbf{k}} B(\mathbf{r}; \mathbf{k})$$

Here, the so-called envelope functions $B(\mathbf{r}; \mathbf{k})$ are periodic in $\mathbf{r}$-space, that is

$$\mathcal{T}_{\mathbf{R}} B(\mathbf{r}; \mathbf{k}) = B(\mathbf{r}; \mathbf{k})$$

We respectively refer $A(\mathbf{r}; \mathbf{k})$ and $B(\mathbf{r}; \mathbf{k})$ to as the wave and envelope functions.

Referring to the above, we may see that both eigenvalues $\mu$ and $\lambda$ are actually functions of $\mathbf{k}$ as $\mu = \mu(\mathbf{k})$ and $\lambda = \lambda(\mathbf{k})$. This leads us to the fact that

$$\mathcal{L} A(\mathbf{r}; \mathbf{k}) = \lambda(\mathbf{k}) A(\mathbf{r}; \mathbf{k})$$

### 2.1. Extensions to the Bloch Theorem

The eigenvalues $\lambda(\mathbf{k})$ turnout to be multi-valued periodic functions in the reciprocal space, so that [2-4]

$$\mathcal{T}_{\mathbf{K}} \lambda(\mathbf{k}) = \lambda(\mathbf{k})$$

A unit-cell of the $\mathbf{k}$-space which constitutes the periodicity of $\lambda(\mathbf{k})$ is known as the Brillouin Zone. Evidently, this unit-cell is extended in the $\mathbf{k}$-space across the basis vectors $\mathbf{b}_1$, $\mathbf{b}_2$, and $\mathbf{b}_3$.

For every given vector $\mathbf{k}$, there are infinitely many of eigenvalue functions $\lambda(\mathbf{k})$ in general. So to be precise, one would need to consider $\lambda(\mathbf{k}) = \lambda_n(\mathbf{k})$, with $n$ being a natural number referred to as the band index. This is exactly what truly happens for the case of every periodic media, such as photonic and plasmonic crystals [6,7,13], electronic crystals [2-4,9-12], and even phononic crystals [14].

If we take advantage of the periodicity of these eigenvalues, we may express $\lambda(\mathbf{k})$ always as $\lambda_n(\mathbf{k}) = \lambda_n(\mathbf{k} - \mathbf{K})$, in such a way that $\mathbf{k} - \mathbf{K}$ would belong to the first Brillouin Zone. This notation with the aid of band-index $n$ makes $\lambda_n(\mathbf{k})$ a single-valued function of its argument. Similarly, as long as $\mathbf{k}$ were restricted to a single Brillouin Zone, then the band index integer $n$ would be needed to resolve the eigenfunctions $A_n(\mathbf{r}; \mathbf{k})$ corresponding to the eigenvalues $\lambda_n(\mathbf{k})$ at equivalent $\mathbf{k}$ points, where any two $\mathbf{k}_1$ and $\mathbf{k}_2$ points are said to be equivalent if $\mathbf{k}_1 - \mathbf{k}_2 = \mathbf{K}$ with $\mathbf{K}$ being a discrete reciprocal lattice vector. From this point on, we drop the explicit dependence of eigenfunctions on the band index $n$ for the sake of brevity, unless needed.

## 2.2. Further Properties of the Bloch Theorem

The above relationships when put together with the basic Bloch theorem, we obtain unexpected result for the first time

$$B(\mathbf{r}; \mathbf{k} + \mathbf{K}) = e^{+j\mathbf{r} \cdot \mathbf{K}} B(\mathbf{r}; \mathbf{k})$$

that is, $B(\mathbf{r}; \mathbf{k})$ must be pseudo-periodic in the $\mathbf{k}$-space, too. This is to be compared to the pseudo-periodicity of $A(\mathbf{r}; \mathbf{k})$ in the $\mathbf{r}$-space.

When these two properties are combined, we arrive at the following conjugate results

$$A(\mathbf{r} + \mathbf{R}; \mathbf{k} + \mathbf{K}) = e^{-j\mathbf{R} \cdot \mathbf{k}} A(\mathbf{r}; \mathbf{k})$$
$$B(\mathbf{r} + \mathbf{R}; \mathbf{k} + \mathbf{K}) = e^{+j\mathbf{r} \cdot \mathbf{K}} B(\mathbf{r}; \mathbf{k})$$

These state that both of the wave $A(\mathbf{r}; \mathbf{k})$ and envelope $B(\mathbf{r}; \mathbf{k})$ functions are biperiodic, however, the wave $A(\mathbf{r}; \mathbf{k})$ is pseudoperiodic in $\mathbf{r}$-space and periodic in $\mathbf{k}$-space, while conversely the envelope $B(\mathbf{r}; \mathbf{k})$ is periodic in $\mathbf{r}$-space and pseudoperiodic in $\mathbf{k}$-space.

These properties furthermore allow a trivial translational-invariance in phase, such as the replacements

$$A(\mathbf{r}; \mathbf{k}) \to e^{-j\theta(\mathbf{r}; \mathbf{k})} A(\mathbf{r}; \mathbf{k})$$
$$B(\mathbf{r}; \mathbf{k}) \to e^{+j\theta(\mathbf{r}; \mathbf{k})} B(\mathbf{r}; \mathbf{k})$$

would provide another equivalent set of eigenfunctions, as long as $\theta(\mathbf{r}; \mathbf{k})$ are real-valued and double-periodic as

$$\mathcal{T}_\mathbf{R} \mathcal{T}_\mathbf{K} \theta(\mathbf{r}; \mathbf{k}) = \theta(\mathbf{r}; \mathbf{k})$$

Choice of $\theta(\mathbf{r}; \mathbf{k})$ is non-trivial for the optimum generation of Wannier functions, to be discussed later in Section 4.

## 3. Orthogonality Relations

Now, both of the wave and envelope functions should satisfy orthogonality relationships in $\mathbf{r}$-space and $\mathbf{k}$-space respectively as

$$\langle A(\mathbf{r};\mathbf{k})|A(\mathbf{r};\mathbf{k}')\rangle_{\mathbf{r}} = a_{\mathbf{k}}\delta(\mathbf{k}-\mathbf{k}')$$
$$\langle B(\mathbf{r};\mathbf{k})|B(\mathbf{r}';\mathbf{k})\rangle_{\mathbf{k}} = b_{\mathbf{r}}\delta(\mathbf{r}-\mathbf{r}')$$

with $a_{\mathbf{k}}$ and $b_{\mathbf{r}}$ being some normalization constants. The inner products are here defined as

$$\langle f(\mathbf{r})|g(\mathbf{r})\rangle_{\mathbf{r}} = \iiint f^*(\mathbf{r})g(\mathbf{r})d^3r$$
$$\langle d(\mathbf{k})|e(\mathbf{k})\rangle_{\mathbf{k}} = \iiint d^*(\mathbf{k})e(\mathbf{k})d^3k$$

After normalization and expansion of the inner products, we get

$$\langle A(\mathbf{r};\mathbf{k})|A(\mathbf{r};\mathbf{k}')\rangle_{\mathbf{r}} = V\delta(\mathbf{k}-\mathbf{k}')$$
$$\langle B(\mathbf{r};\mathbf{k})|B(\mathbf{r}';\mathbf{k})\rangle_{\mathbf{k}} = U\delta(\mathbf{r}-\mathbf{r}')$$

While the first relation is simply known as the orthogonality of wave functions, the second one actually is an expression of the completeness relationship.

In the reduced-zone scheme, we may rewrite and renormalize the above equations as

$$\langle A_n(\mathbf{r};\mathbf{k})|A_{n'}(\mathbf{r};\mathbf{k}')\rangle_{\mathbf{r}-\text{Cell}} = V\delta_{nn'}\delta_{\mathbf{kk}'}$$
$$\langle B_m(\mathbf{r};\mathbf{k})|B_{m'}(\mathbf{r}';\mathbf{k})\rangle_{\mathbf{k}-\text{Cell}} = U\delta_{mm'}\delta_{\mathbf{rr}'}$$

where the integration of inner products are restricted to unit cells as

$$\langle f(\mathbf{r})|g(\mathbf{r})\rangle_{\mathbf{r}-\text{Cell}} = \iiint_{\mathbf{r}-\text{Cell}} f^*(\mathbf{r})g(\mathbf{r})d^3r$$

$$\langle d(\mathbf{k})|e(\mathbf{k})\rangle_{\mathbf{k}-\text{Cell}} = \iiint_{\mathbf{k}-\text{Cell}} d^*(\mathbf{k})e(\mathbf{k})d^3k$$

Here, $\mathbf{r}-\text{Cell}$ and $\mathbf{k}-\text{Cell}$ refer actually to the lattice's Unit Cell and Brillouin Zone, respectively.

### 4. Modified Wannier Functions

The orthogonality relationships provide us with two sets of related, yet different, Wannier functions [9,13,15-18] in in $\mathbf{r}$-space and $\mathbf{k}$-space, respectively defined in the reduced zone-schemes as

$$Y_{n\mathbf{R}}(\mathbf{r}) = \frac{1}{\sqrt{U}}\iiint_{\mathbf{k}-\text{Cell}} e^{+j\mathbf{k}\cdot\mathbf{R}}A_n(\mathbf{r};\mathbf{k})d^3k$$

$$Z_{m\mathbf{K}}(\mathbf{k}) = \frac{1}{\sqrt{V}}\iiint_{\mathbf{r}-\text{Cell}} e^{-j\mathbf{K}\cdot\mathbf{r}}B_m(\mathbf{r};\mathbf{k})d^3r$$

As a result, we obtain the shift-properties $Y_{n\mathbf{R}}(\mathbf{r}) = Y_{n\mathbf{0}}(\mathbf{r} - \mathbf{R})$ and $Z_{m\mathbf{K}}(\mathbf{k}) = Z_{m\mathbf{0}}(\mathbf{k} - \mathbf{K})$. This allows us to ignore the discrete vector indices, and simplify the definitions as

$$Y_n(\mathbf{r}) = \frac{1}{\sqrt{U}} \iiint_{\mathbf{k}-\text{Cell}} A_n(\mathbf{r}; \mathbf{k}) d^3k$$

$$Z_m(\mathbf{k}) = \frac{1}{\sqrt{V}} \iiint_{\mathbf{r}-\text{Cell}} B_m(\mathbf{r}; \mathbf{k}) d^3r$$

where appropriate. Wannier functions now turn out to be readily orthogonal as

$$\langle Y_{n\mathbf{R}} | Y_{n'\mathbf{R}'} \rangle_{\mathbf{r}} = \delta_{nn'} \delta_{\mathbf{RR}'}$$
$$\langle Z_{m\mathbf{K}} | Z_{m'\mathbf{K}'} \rangle_{\mathbf{k}} = \delta_{mm'} \delta_{\mathbf{KK}'}$$

Finally, the original wave and envelope functions may be reconstructed from the respective Wannier functions through discrete summations over the lattice points and reciprocal lattice points as

$$A_n(\mathbf{r}; \mathbf{k}) = \frac{(2\pi)^3}{\sqrt{U}} \sum_{\mathbf{R}} e^{-j\mathbf{k}\cdot\mathbf{R}} Y_{n\mathbf{R}}(\mathbf{r})$$

$$B_m(\mathbf{r}; \mathbf{k}) = \frac{(2\pi)^3}{\sqrt{V}} \sum_{\mathbf{K}} e^{+j\mathbf{K}\cdot\mathbf{r}} Z_{m\mathbf{K}}(\mathbf{k})$$

Alternatively, we may rewrite the above definitions as

$$Y_{n\mathbf{R}}(\mathbf{r}) = \frac{1}{\sqrt{U}} \iiint_{\mathbf{k}-\text{Cell}} e^{+j\mathbf{k}\cdot(\mathbf{R}-\mathbf{r})} B_n(\mathbf{r}; \mathbf{k}) d^3k$$

$$Z_{m\mathbf{K}}(\mathbf{k}) = \frac{1}{\sqrt{V}} \iiint_{\mathbf{r}-\text{Cell}} e^{-j(\mathbf{K}-\mathbf{k})\cdot\mathbf{r}} A_m(\mathbf{r}; \mathbf{k}) d^3r$$

This allows us to obtain the direct transformation pairs between these Wannier functions as

$$Y_{n\mathbf{R}}(\mathbf{r}) = (2\pi)^{\frac{3}{2}} \sum_{\mathbf{K}} e^{+j\mathbf{K}\cdot\mathbf{r}} \iiint_{\mathbf{k}-\text{Cell}} e^{+j\mathbf{k}\cdot(\mathbf{R}-\mathbf{r})} Z_{n\mathbf{K}}(\mathbf{k}) d^3k$$

$$Z_{m\mathbf{K}}(\mathbf{k}) = (2\pi)^{\frac{3}{2}} \sum_{\mathbf{R}} e^{-j\mathbf{k}\cdot\mathbf{R}} \iiint_{\mathbf{r}-\text{Cell}} e^{-j(\mathbf{K}-\mathbf{k})\cdot\mathbf{r}} Y_{m\mathbf{R}}(\mathbf{r}) d^3r$$

Here, we note the use of identical band indices on both sides.

It should be mentioned that the translationally-invariant phase $\theta(\mathbf{r}; \mathbf{k})$ takes significant part in the integration over eigenfunctions, which may be illustrated explicitly as

$$Y_n(\mathbf{r}) = \frac{1}{\sqrt{U}} \iiint_{\mathbf{k}-\text{Cell}} e^{-j\theta(\mathbf{r};\mathbf{k})} A_n(\mathbf{r};\mathbf{k}) d^3k$$

$$Z_m(\mathbf{k}) = \frac{1}{\sqrt{V}} \iiint_{\mathbf{r}-\text{Cell}} e^{+j\theta(\mathbf{r};\mathbf{k})} B_m(\mathbf{r};\mathbf{k}) d^3r$$

Since there are infinitely many choices for $\theta(\mathbf{r};\mathbf{k})$, Wannier functions would not be unique, and normally need to be constructed in such a way to be maximally-localized. This is normally done through global minimization of a spread functional; the reader is referred to literature for in-depth discussion of this issue [15], [17-20], [21], [9,22] for plasmonic, photonic, phononic, and electronic crystals, respectively.

## 5. Example

### 5.1. Two-dimensional Photonic Crystals

Consider a two-dimensional (2D) photonic crystal (PC) with in-plane propagation, which allows separation of E-polarized and H-polarized modes. Here, the 2D relative permittivity and permeability function makes the whole structure periodic as $\epsilon_r(\mathbf{r}) = \epsilon_r(\mathbf{r}+\mathbf{R})$ and $\mu_r(\mathbf{r}) = \mu_r(\mathbf{r}+\mathbf{R})$ with $\mathbf{R} = n_1\mathbf{a}_1 + n_2\mathbf{a}_2$. The operator for the normal component of the electric field $A(\mathbf{r},\mathbf{k}) = \sqrt{\epsilon_r(\mathbf{r})} E_z(\mathbf{r},\mathbf{k}) = \sqrt{\epsilon_r(\mathbf{r})} e^{-j\mathbf{k}\cdot\mathbf{r}} F_z(\mathbf{r},\mathbf{k})$ would become [6]

$$\mathcal{L}A(\mathbf{r},\mathbf{k}) = \lambda A(\mathbf{r},\mathbf{k})$$

The scalar operator and eigenvalue is in the extended zone now given as

$$\mathcal{L}_E^{(2)} = \frac{1}{\sqrt{\epsilon_r(\mathbf{r})}} \nabla \cdot \left\{ \frac{1}{\mu_r(\mathbf{r})} \nabla \left[ \frac{1}{\sqrt{\epsilon_r(\mathbf{r})}} (\cdot) \right] \right\}$$

$$\lambda(\mathbf{k}) = \frac{\omega_E^2(\mathbf{k})}{c^2}$$

with $\nabla = \frac{\partial}{\partial x}\hat{x} + \frac{\partial}{\partial y}\hat{y}$. Furthermore, $\omega_E(\mathbf{k})$ is the E-polarization band structure, and $\mathcal{L}_E^{(2)}$ is easy to be shown to be self-adjoint operator under these criteria, hence the eigenfunctions obeying the orthogonality relationships after corresponding displaying the band indices

$$\langle A(\mathbf{r};\mathbf{k}) | A(\mathbf{r};\mathbf{k}') \rangle_{\mathbf{r}} = \iint \epsilon_r(\mathbf{r}) E_{nz}(\mathbf{r};\mathbf{k}) E_{n'z}^*(\mathbf{r};\mathbf{k}') d^2r = V\delta_{nn'}\delta(\mathbf{k}-\mathbf{k}')$$

$$\langle B(\mathbf{r};\mathbf{k}) | B(\mathbf{r}';\mathbf{k}) \rangle_{\mathbf{k}} = \iint \epsilon_r(\mathbf{r}) F_{mz}(\mathbf{r};\mathbf{k}) F_{m'z}^*(\mathbf{r}';\mathbf{k}) d^2k = U\delta_{mm'}\delta(\mathbf{r}-\mathbf{r}')$$

where $V$ actually denotes the area of the unit-cell; for this to happen, one may simply set $\mathbf{a}_3 = \hat{z}$. Note that, we have $VU = (2\pi)^2$ for the 2D PC. The relevant pair of E-polarization Wannier functions [19,20] would be

$$Y_n(\mathbf{r}) = \frac{\sqrt{\epsilon_r(\mathbf{r})}}{\sqrt{U}} \iint_{\mathbf{k}-\text{Cell}} E_{nz}(\mathbf{r};\mathbf{k})d^2k$$

$$Z_m(\mathbf{k}) = \frac{1}{\sqrt{V}} \iint_{\mathbf{r}-\text{Cell}} \sqrt{\epsilon_r(\mathbf{r})} F_{mz}(\mathbf{r};\mathbf{k})d^2r$$

Similarly, the relationships of the H-polarized components could be found. As for the scalar operator of the normal component of the magnetic field $A(\mathbf{r},\mathbf{k}) = \sqrt{\mu_r(\mathbf{r})}H_z(\mathbf{r},\mathbf{k}) = \sqrt{\mu_r(\mathbf{r})}e^{-j\mathbf{r}\cdot\mathbf{k}}G_z(\mathbf{r},\mathbf{k})$, one would obtain

$$\mathcal{L}_H^{(2)} = \frac{1}{\sqrt{\mu_r(\mathbf{r})}} \nabla \cdot \left\{ \frac{1}{\epsilon_r(\mathbf{r})} \nabla \left[ \frac{1}{\sqrt{\mu_r(\mathbf{r})}} (\cdot) \right] \right\}$$

$$\lambda(\mathbf{k}) = \frac{\omega_H^2(\mathbf{k})}{c^2}$$

$\omega_H(\mathbf{k})$ is the H-polarization band structure, and $\mathcal{L}_H^{(2)}$ is self-adjoint. Hence the eigenfunctions obey the orthogonality relationships after corresponding displaying the band indices

$$\langle A(\mathbf{r};\mathbf{k})|A(\mathbf{r};\mathbf{k}')\rangle_\mathbf{r} = \iint \mu_r(\mathbf{r})H_{nz}(\mathbf{r};\mathbf{k})H_{n'z}^*(\mathbf{r};\mathbf{k}')d^2r = V\delta_{nn'}\delta(\mathbf{k}-\mathbf{k}')$$

$$\langle B(\mathbf{r};\mathbf{k})|B(\mathbf{r}';\mathbf{k})\rangle_\mathbf{k} = \iint \mu_r(\mathbf{r})G_{mz}(\mathbf{r};\mathbf{k})G_{m'z}^*(\mathbf{r}';\mathbf{k})d^2k = U\delta_{mm'}\delta(\mathbf{r}-\mathbf{r}')$$

where $V$ actually denotes the area of the unit-cell; for this to happen, one may simply set $\mathbf{a}_3 = \hat{z}$. Similarly, we have $VU = (2\pi)^2$ for the 2D PC. The relevant pair of H-polarization Wannier functions would be

$$Y_n(\mathbf{r}) = \frac{\sqrt{\mu_r(\mathbf{r})}}{\sqrt{U}} \iint_{\mathbf{k}-\text{Cell}} H_{nz}(\mathbf{r};\mathbf{k})d^2k$$

$$Z_m(\mathbf{k}) = \frac{1}{\sqrt{V}} \iint_{\mathbf{r}-\text{Cell}} \sqrt{\mu_r(\mathbf{r})} G_{mz}(\mathbf{r};\mathbf{k})d^2r$$

## 5.2. Three-dimensional Photonic Crystals

If the permittivity and permeability functions are three-dimensional (3D) periodic tensor of space, then within the frame of reference which $[\epsilon_r(\mathbf{r})]$ becomes diagonal (under no loss and no optical activity [9-11]), one would have the following relation for the vector operator

$$\mathcal{L}_E^{(3)} = -\frac{1}{\sqrt{[\epsilon_r(\mathbf{r})]}} \nabla \times \left\{ \frac{1}{[\mu_r(\mathbf{r})]} \nabla \times \left[ \frac{1}{\sqrt{[\epsilon_r(\mathbf{r})]}} (\cdot) \right] \right\}$$

where $\sqrt{[\epsilon_r(\mathbf{r})]} \equiv [\epsilon_{r\,mm}^{1/2}(\mathbf{r})\delta_{mn}]$ is properly defined because of its diagonal form. Similarly, we obtain

$$\mathcal{L}_E^{(3)}\mathbf{E}_n(\mathbf{r},\mathbf{k}) = \frac{\omega_n^2(\mathbf{k})}{c^2}\mathbf{E}_n(\mathbf{r},\mathbf{k})$$

which leads to the orthogonality relations

$$\langle A(\mathbf{r};\mathbf{k})|A(\mathbf{r};\mathbf{k}')\rangle_\mathbf{r} = \iiint \left\{\sqrt{[\epsilon_r(\mathbf{r})]}\mathbf{E}_n(\mathbf{r},\mathbf{k})\right\} \otimes \left\{\sqrt{[\epsilon_r(\mathbf{r})]}\mathbf{E}_{n'}^{\,*}(\mathbf{r};\mathbf{k}')\right\} d^3r = V\delta_{nn'}\delta(\mathbf{k}-\mathbf{k}')[I]$$

$$\langle B(\mathbf{r};\mathbf{k})|B(\mathbf{r}';\mathbf{k})\rangle_\mathbf{k} = \iiint \left\{\sqrt{[\epsilon_r(\mathbf{r})]}\mathbf{F}_m(\mathbf{r},\mathbf{k})\right\} \otimes \left\{\sqrt{[\epsilon_r(\mathbf{r})]}\mathbf{F}_{m'}^{\,*}(\mathbf{r}';\mathbf{k})\right\} d^3k = U\delta_{mm'}\delta(\mathbf{r}-\mathbf{r}')[I]$$

Here, $\otimes$ represents the tensor outer product, and $[I]$ is the $3 \times 3$ unit tensor. From these equations the additional forms are inferred by taking the trace

$$\iiint \mathrm{tr}\left(\left\{\sqrt{[\epsilon_r(\mathbf{r})]}\mathbf{E}_n(\mathbf{r},\mathbf{k})\right\} \otimes \left\{\sqrt{[\epsilon_r(\mathbf{r})]}\mathbf{E}_{n'}^{\,*}(\mathbf{r};\mathbf{k}')\right\}\right) d^3r = 3V\delta_{nn'}\delta(\mathbf{k}-\mathbf{k}')$$

$$\iiint \mathrm{tr}\left(\left\{\sqrt{[\epsilon_r(\mathbf{r})]}\mathbf{F}_m(\mathbf{r},\mathbf{k})\right\} \otimes \left\{\sqrt{[\epsilon_r(\mathbf{r})]}\mathbf{F}_{m'}^{\,*}(\mathbf{r}';\mathbf{k})\right\}\right) d^3k = 3U\delta_{mm'}\delta(\mathbf{r}-\mathbf{r}')$$

Finally, when the dielectric is isotropic, one would reach the scalar equations

$$\iiint \epsilon_r(\mathbf{r})\mathbf{E}_n(\mathbf{r},\mathbf{k}) \cdot \mathbf{E}_{n'}^{\,*}(\mathbf{r};\mathbf{k}')d^3r = 3V\delta_{nn'}\delta(\mathbf{k}-\mathbf{k}')$$

$$\iiint \epsilon_r(\mathbf{r})\mathbf{F}_m(\mathbf{r},\mathbf{k}) \cdot \mathbf{F}_{m'}^{\,*}(\mathbf{r}';\mathbf{k})d^3k = 3U\delta_{mm'}\delta(\mathbf{r}-\mathbf{r}')$$

The relevant pair of vector E-polarization Wannier functions would be obtainable from generalization of the vector Wannier functions [19]

$$\mathbf{Y}_n(\mathbf{r}) = \frac{\sqrt{[\epsilon_r(\mathbf{r})]}}{\sqrt{U}} \iiint_{\mathbf{k}-\mathrm{Cell}} \mathbf{E}_{nz}(\mathbf{r};\mathbf{k})d^3k$$

$$\mathbf{Z}_m(\mathbf{k}) = \frac{1}{\sqrt{V}} \iiint_{\mathbf{r}-\mathrm{Cell}} \sqrt{[\epsilon_r(\mathbf{r})]}\mathbf{F}_{mz}(\mathbf{r};\mathbf{k})d^3r$$

Identical deductions may be made for the magnetic field eigenfunctions $\mathbf{H}(\mathbf{r},\mathbf{k}) = e^{-j\mathbf{r}\cdot\mathbf{k}}\mathbf{G}(\mathbf{r},\mathbf{k})$ once $[\mu_r(\mathbf{r})]$ is diagonalized, which may be found by interchanging permittivity and permeability, as well as Electric and Magnetic fields everywhere.

## 6. Conclusions

In this article, we have presented an in-depth discussion on the basic properties of Bloch waves in periodic media, and obtained a novel biperiodicity property in the eigenfunctions for the first time. We showed that the envelope and wave functions both satisfy mutual periodic and pseudoperiodic properties in the physical and reciprocal spaces, which shows there exist much more similarity between these two spaces than thought before. Extensions of these property were found to be applicable to defining a new set of Wannier functions in reciprocal space, rather than the conventional definition in physical space, while these two Wannier functions are directly connected through unorthodox transformations. It is envisioned that the future investigation of these pair of Wannier functions would result in more efficient analytical tools of periodic media, and simplify the construction of maximally-localized band functions.

**Appendix A**

These identities are useful in calculation of Fourier series and expansion terms of periodic functions

$$\mathcal{T}_\mathbf{R} f(\mathbf{r}) = f(\mathbf{r}) = \frac{1}{(2\pi)^3} \sum_\mathbf{K} f_\mathbf{K} e^{-j\mathbf{K}\cdot\mathbf{r}}$$

$$f_\mathbf{K} = \frac{1}{V} \iiint_{\mathbf{r}-\text{Cell}} e^{+j\mathbf{K}\cdot\mathbf{r}} f(\mathbf{r}) d^3 r$$

and

$$\mathcal{T}_\mathbf{K} g(\mathbf{k}) = g(\mathbf{k}) = \frac{1}{(2\pi)^3} \sum_\mathbf{R} g_\mathbf{R} e^{+j\mathbf{R}\cdot\mathbf{r}}$$

$$g_\mathbf{R} = \frac{1}{U} \iiint_{\mathbf{k}-\text{Cell}} e^{-j\mathbf{k}\cdot\mathbf{R}} g(\mathbf{k}) d^3 k$$

for $\mathbf{R} = n_1 \mathbf{a}_1 + n_2 \mathbf{a}_2 + n_3 \mathbf{a}_3$ and $\mathbf{K} = m_1 \mathbf{b}_1 + m_2 \mathbf{b}_2 + m_3 \mathbf{b}_3$.

**Appendix B**

Self-adjointness property of $\mathcal{L} = \mathcal{L}^\dagger$ directly leads to an orthogonality relation for wave functions $A(\mathbf{r}; \mathbf{k})$ in the **r**-space. However, we may make the transformation

$$\mathcal{K}_\mathbf{k} = \frac{1}{\lambda(\mathbf{k})} e^{+j\mathbf{k}\cdot\mathbf{r}} \mathcal{L} e^{-j\mathbf{k}\cdot\mathbf{r}}$$

which allows us to write

$$\mathcal{K}_\mathbf{k} B(\mathbf{r}; \mathbf{k}) = B(\mathbf{r}; \mathbf{k})$$

being subject to periodic boundary conditions across **r**-cells. The new operator $\mathcal{K}_\mathbf{k}$ is not necessarily self-adjoint, and hence the set of envelope functions $B(\mathbf{r}; \mathbf{k})$ are not necessarily orthogonal in the **r**-space. If $\mathcal{L}$ is expressible in terms of the gradient operator $\nabla$, then $\mathcal{K}_\mathbf{k}$ is obtainable from $\mathcal{L}$ simply by the replacement $\nabla \rightarrow \nabla - j\mathbf{k}$.

Now, noting that $\lambda(\mathbf{k})$ must be periodic in **k**-space, the simultaneous pair of transformations $\mathcal{K}_\mathbf{k} \rightarrow \mathcal{K}_{\mathbf{k}+\mathbf{K}}$ and $B(\mathbf{r}; \mathbf{k}) \rightarrow B(\mathbf{r}; \mathbf{k}+\mathbf{K})$ easily result in the desired pseudo-periodicity of $B(\mathbf{r}; \mathbf{k})$ in **k**-space as described in Section 3.

## Appendix C

Now, we define the pair of Fourier transformations

$$\mathbb{F}\{f(\mathbf{r})\}(\boldsymbol{\kappa}) = \frac{1}{(2\pi)^{\frac{3}{2}}} \iiint e^{+j\boldsymbol{\kappa}\cdot\mathbf{r}} f(\mathbf{r}) d^3r$$

$$\mathbb{F}^{-1}\{g(\mathbf{k})\}(\boldsymbol{\rho}) = \frac{1}{(2\pi)^{\frac{3}{2}}} \iiint e^{-j\mathbf{k}\cdot\boldsymbol{\rho}} g(\mathbf{k}) d^3k$$

and perform Fourier transformations on both arguments as and mapping $(\mathbf{r}, \mathbf{k})$ first to $(\boldsymbol{\kappa}, \boldsymbol{\rho})$, then renaming $\boldsymbol{\kappa} \to \mathbf{k}$ and $\boldsymbol{\rho} \to \mathbf{r}$, to get back to the space $(\mathbf{r}, \mathbf{k})$. But, this time we reach a different yet related eigenvalue problem

$$\mathcal{J} C(\mathbf{r}; \mathbf{k}) = \nu(\mathbf{r}) C(\mathbf{r}; \mathbf{k})$$

in which $\mathcal{J}$ is self-adjoint, too, given by

$$\mathcal{J} = \mathcal{J}^\dagger$$

Ideally, in order to construct the operator $\mathcal{J}$, one would need to make the replacements $\mathbf{r} \to -j\,\partial/\partial\mathbf{k}$ and $\mathbf{k} \to +j\,\partial/\partial\mathbf{r}$, albeit simultaneously. Unfortunately, this is not typically doable in most practical problems, since the extensive algebraic form of $\lambda(\mathbf{k})$ must be known.

## Acknowledgement

The author is indebted to Dr. Ali Naqavi at the École polytechnique fédérale de Lausanne for reading the initial manuscript and providing insightful comments.